\documentclass[12pt]{iopart}

\begin{document}

\title[Variations on a theme of q-oscillator]{Variations on a theme of q-oscillator}

\author{Oktay K. Pashaev}

\address{Izmir Institute of Technology, Gulbahce Campus, Urla, Izmir 35430, Turkey}
\ead{oktaypashaev@iyte.edu.tr}
\vspace{10pt}
\begin{indented}
\item[]October  2014
\end{indented}

\begin{abstract}
  We present several ideas in direction of physical interpretation of $q$- and $f$-oscillators as a nonlinear oscillators.
  First we show that an arbitrary one dimensional integrable system in action-angle variables can be naturally represented as a classical and quantum $f$-oscillator.
 As an example, the semi-relativistic oscillator as a descriptive of the Landau levels for relativistic electron in magnetic field is solved as an $f$-oscillator.  By using dispersion relation
 for $q$-oscillator we solve the linear q-Schr\"odinger equation and corresponding nonlinear complex q-Burgers equation.
 The same dispersion allows us to construct integrable q-NLS model as a deformation of cubic NLS in terms of recursion operator of NLS hierarchy.
 Peculiar property of the model is to be completely integrable at any order of expansion in deformation parameter around $q=1$.
 As another variation on the theme, we consider hydrodynamic flow in bounded domain. For the flow bounded by two concentric circles we formulate the two circle theorem and construct solution as the q-periodic flow by non-symmetric $q$-calculus.
 Then we generalize this theorem to the flow in the wedge domain bounded by two arcs. This two circular-wedge theorem determines images of the flow  by extension
 of $q$-calculus to two bases: the real one, corresponding to circular arcs and the complex one, with $q$ as a primitive root of unity.
 As an application, the vortex motion in annular domain as a nonlinear oscillator in the form of classical and quantum f-oscillator is studied.
Extending idea of q-oscillator to two bases with the golden ratio, we describe Fibonacci numbers as a special type of $q$-numbers with matrix Binet formula. We derive the corresponding  golden quantum oscillator, nonlinear coherent states and Fock-Bargman representation. The spectrum of it satisfies the triple relations, while the energy levels relative difference approaches asymptotically to the golden ratio and has no classical limit.
\end{abstract}


\vspace{2pc}

\section{Introduction}

The harmonic oscillator plays the central role in classical and modern science, starting from elementary pendulum in XVII century, when ideas of Johannes  Kepler signify important intermediate step between previous magic-symbolical and modern quantitative-mathematical description of nature \cite{Pauli1}.
At that time the birth of classical physics was caused by application of an abstract idea of periodicity to concrete variety of problems \cite{Whitehead}.
Wave theory and Fourier series are some tools developed from this concept in classical physics.
But only in modern science we observe essentially new qualitative transition from dynamics to the oscillation theory. It was emphasized by L. Mandelstam in lectures on
oscillation theory \cite{Mandelstam},  that the difference between standard dynamics and the oscillation theory is that in dynamics we are interested in description of what is going on in a given place at a given time. While in the oscillation theory - in the motion of the system as a whole. In classical mechanical picture of the world,  positions and velocities are primary objects, while oscillations are the secondary ones.  However, starting from quantum mechanics, in which our knowledge of simultaneous positions and velocities is restricted by the Heisenberg  uncertainty relations, this point of view has been drastically changed. The wave mechanics affirms that the wholeness of the quantum process is something of the same primacy as position of a particle. And every particle is associated with some stationary oscillation process.
Essential to note that knowledge of phase of the particle and knowledge of the stationary state are excluding each other. The phase of the particle (trajectory) in a single stationary state does not exist, since any attempt to derive this phase switches the system to another stationary state \cite{Pauli2}.
This way the basic characteristic of oscillation theory, as a consideration of the process as a whole, becomes underlie of fundamental questions of mechanics in quantum world. This point of view also influenced the modern theory of nonlinear dynamical systems, studying the qualitative behavior of a system in the phase space, its integrability properties and chaos \cite{Andronov}.

In recent development of quantum integrable systems the concept of quantum group as a deformation of the Lie group with deformation parameter $q$  was discovered. The notion of quantum $q$-oscillator  as a $q$-deformed harmonic oscillator was introduced in studies on quantum Heisenberg-Weyl group \cite{Biedenharn}, \cite{Macfarlane}, \cite{Sun}. This oscillator is related with so called
symmetric $q$-calculus \cite{Kac}, while another version of $q$-oscillator \cite{Arik}, \cite{Kuryshkin} with non-symmetric $q$-calculus. The difference is in definition of $q$-number and several generalizations of these numbers and oscillators with different basis were found.
 In the large stream of articles devoted to $q$-oscillator here we like to emphasize the set of papers published by V. I. Man'ko and coauthors, which are not widely known \cite{Manko1}, \cite{Manko2}, \cite{Manko3}.

In  these papers the physical approach to q-oscillator as a nonlinear oscillator was developed. It was shown that it is an oscillator with frequency depending on its energy in the form of hyperbolic cosine function of the energy. The classical motion of this nonlinear oscillator becomes descriptive of the motion of a q-oscillator and the frequency of oscillations is increasing exponentially with energy. By standard quantization of this nonlinear system, the authors got a quantum q-oscillator as a nonlinear quantum oscillator with anharmonicity described by power series in energy. Then they generalized the approach to an arbitrary energy dependence of frequency and called it as the $f$-oscillator. In fact, in addition to frequency, any constant parameters in some exactly solvable system can be replaced by integrals of motion. This idea in some sense continues Dirac's approach to fundamental constants as a simple functions slowly varying in time. The slow variation of parameters can be implemented by adiabatic invariants and integrals of motion, with corresponding quantization. Then starting from any integrable system one can replace parameters of the system by integrals of motion and get the hierarchy of integrable systems from the given one.

In the present paper we develop several ideas as a variations on such approach to $q$- and $f$-oscillators as the main theme. In Section 2 we show that an arbitrary one dimensional integrable model in action-angle variables has natural description as a classical and quantum $f$-oscillator. As an example we consider semi-relativistic oscillator, related with Landau levels for relativistic electron in magnetic field. In Section 3 we review some results on symmetric $q$-oscillator and in the next Section 4 we study linear q-Schrodinger equation with dispersion relation of this oscillator. The symmetry operators and polynomial solutions with moving zeros are described, as well as nonlinearization of the model by complex Cole-Hopf-Madelung transformation. The symmetry of the linear equation then is rewritten as B\"acklund transformation for the nonlinear q-Burgers equation. In Section 6, following the general procedure developed in \cite{Pashaev3}, we construct the nonlinear Schr\"odinger equation with symmetric $q$-dispersion. This q-NLS equation is an integrable model from NLS hierarchy with Lax representation, infinite number of integrals of motion, soliton solutions etc. In the limit $q \rightarrow 1$ it reduces to NLS, which is one of the universal soliton equations
and for $q \approx 1$ provide higher order corrections in dispersion and nonlinearity. Peculiar property of the model is to be completely integrable at any order of expansion in deformation parameter around $q=1$.
As a next variation on q-calculus application in Section 4 we consider hydrodynamic flow in bounded circular wedge domain. We formulate general two circular wedge theorem and show that the flow is determined  by q-periodic functions. For real $q= r^2_2/r^2_1$ it is the flow in annular concentric circles domain, while for wedge with angle $\pi/n$, $q$ is the primitive root of unity. For a vortex problem in such domains we describe full set of vortex images as a kaleidoscope in terms of $q$-elementary functions. As an application we describe the point vortex motion in annular domain as a nonlinear oscillator with frequency depending on radius of motion. Then we find the $f$-oscillator form of this model and discuss corresponding quantization. In the last Section 10 by introducing matrix form of Binet formula for Fibonacci numbers, we solve so called golden oscillator and find corresponding coherent states and Fock-Bargman type representation. Details of some proofs are given in Appendix.

I apologize that in this paper would be not able to give complete list of references in this very wide field of research. Instead of this I try to represent some basic ideas in a pedagogical way.

\section{Integrable models and nonlinear oscillators}

Representation of an integrable model by the action-angle variables allows one to interpret the model as a set of nonlinear oscillators.
For simplicity here we illustrate the idea only for the one degree of freedom Hamiltonian system. Due to conservation of energy this
system is integrable. Let $H(p,q)$ is Hamiltonian function with canonical variables $(q, p)$. The action and angle variables $(J, \theta)$ are introduced by
generating function, \cite{Sagdeev}
\begin{equation}
S(q, J) = S(q, H(J))= \int^q p(q, H) \rmd q,
\end{equation}
as the abbreviated action, in the following way
\begin{equation}
J = \frac{1}{2\pi} \oint p(q, H) \rmd q = J(H),\,\,\,\,\,\, \theta = \frac{\partial S(q,J)}{\partial J}.
\end{equation}
We suppose that the motion is finite and integral is taken along the full period of oscillations. Hamilton's equations of motion in these
variables are
\begin{equation}
\dot J = - \frac{\partial H(J)}{\partial \theta} = 0,\,\,\,\,\, \dot \theta = \frac{\partial H(J)}{\partial J} \equiv \omega(J).
\end{equation}
The first equation implies that $H$ is function of $J$ only, independent of $\theta$ (cyclic coordinate), and the action variable,  as well as the energy $E$ is an integral of motion. The second equation determines nonlinear frequency $\omega = \omega(J)$ and solution
\begin{equation}
J = J(E),\,\,\,\,\,\, \theta(t) = \omega(J) t + \theta_0.
\end{equation}
  The change of $S$ and $\theta$ in the period is
 \begin{equation}
 \Delta S = \oint p \rmd q = 2 \pi J,\,\,\,\,\Delta \theta = \frac{\partial \Delta S}{\partial J} = 2 \pi,
 \end{equation}
and  trajectory of the system represents a curve on the cylindrical surface in phase space $(p, q)$ endowed with time axis $t$ \cite{Sagdeev}.
 Then, original variable $(q, p)$ as functions of $(J, \theta)$ are periodic and can be decomposed to Fourier series as the spectral
 decomposition
 \begin{eqnarray}
 q(J, \theta) = \sum^\infty_{n=-\infty} q_n(J) \rme^{\rmi n \theta}, \,\,\,\,
 p(J, \theta) = \sum^\infty_{n=-\infty} p_n(J) \rme^{\rmi n \theta}.\label{spectraldecomposition}
 \end{eqnarray}

\subsubsection{Linear oscillator}
 As an example, the linear oscillator
 \begin{equation}
 H_0 = \frac{p^2}{2m} + \frac{m \omega^2_0 q^2}{2}
 \end{equation}
 in action variables is
 \begin{equation}
 H_0(J) = \omega_0 J,\,\,\,\,\omega(J) = \omega_0 \label{linearoscillator}
 \end{equation}
 with the following solution 
 \begin{equation}
 J = \frac{H_0}{\omega_0},\,\,\,\theta(t) = \omega_0 t + \theta_0.
 \end{equation}
 The spectral decomposition has only harmonics with $n=\pm1$ and
 \begin{equation}
 q(J, \theta) = \sqrt{\frac{2 J}{m \omega_0}} \sin (\omega_0 t + \theta_0),\,\,\,p(J, \theta) = \sqrt{2 J m \omega_0} \cos (\omega_0 t + \theta_0).
 \end{equation}
 Appearance of other modes in the decomposition (\ref{spectraldecomposition}) means anharmonicity  in oscillations.

\subsubsection{Complex coordinates}
 Quantization of action-angle variables implies that the action variable $J$ is replaced by Hermitian operator $\hat J$, while instead of $\theta$
 one should use the unitary operator $\rme^{\rmi \theta}$ \cite{Pauli2}, \cite{Dirac}. This suggests that proper variables for quantization of the
 system in action-angle variables  $\{ \theta, J  \} = 1$ are complex functions
 \begin{equation}
   \alpha = \rmi\sqrt{J} \rme^{-\rmi\theta},\,\,\,\,\,\,\bar \alpha = -\rmi\sqrt{J} \rme^{\rmi\theta}, \label{complexcoordinate}
   \end{equation}
   with canonical bracket
   \begin{equation}
 \{ \alpha, \bar\alpha  \} = -\rmi.
   \end{equation}
 Since $J = \alpha \bar\alpha$, the Hamiltonian of the linear oscillator (\ref{linearoscillator}) in these variables is just
 $H_0 = \alpha \bar\alpha$ (for simplicity we put $\omega_0 =1$)
 and generic nonlinear Hamiltonian is function of $H_0$:
 \begin{equation}
 H = H(J) = H(\alpha \bar \alpha) = H(H_0). \label{genericH}
 \end{equation}
 It determines evolution equation as a nonlinear oscillator
 \begin{equation}
 \dot\alpha = - \rmi \frac{\partial H(\alpha \bar\alpha)}{\partial \bar\alpha} = -\rmi H'(|\alpha|^2) \alpha = - \rmi \omega(|\alpha|^2) \alpha, \label{evolution1}
 \end{equation}
 with frequency $\omega (|\alpha|^2)$ depending on value of amplitude, where the last one is an integral of motion.

\subsection{Classical f-oscillator}
For generic $H = H(J) = H(\alpha \bar\alpha)$ as in (\ref{genericH}), by introducing complex variables
\begin{equation}
      \alpha_f = \sqrt{\frac{H(J)}{J}} \alpha,\,\,\,\,\,\bar \alpha_f = \sqrt{\frac{H(J)}{J}} \bar\alpha ,\label{ftr}
      \end{equation}
where $J = \bar\alpha \alpha$, we express the model in the form of the f-oscillator \cite{Manko3},
\begin{equation}
H(\alpha_f, \bar \alpha_f) = \alpha_f \bar \alpha_f,
\end{equation}
with the Poisson bracket
\begin{equation}
\{ \alpha_f, \bar \alpha_f  \} = - \rmi \frac{\partial H}{\partial J}(\alpha_f, \bar \alpha_f) = - \rmi \omega(\alpha_f, \bar \alpha_f)
\end{equation}
and evolution determined by the same frequency as in (\ref{evolution1})
\begin{equation}
\dot \alpha_f = - \rmi \omega \alpha_f.
\end{equation}
\subsection{Quantum f-oscillator}
Quantization of this system replaces complex variables $\alpha$, $\bar\alpha$ with bosonic operators $[a, a^+] = 1$, $N = a^+ a$.
Then the Hamiltonian of the corresponding quantum f- oscillator is \cite{Manko3},
\begin{equation}
H = \frac{a_f a_f^+ + a_f^+ a_f}{2},\label{fHamilton}
\end{equation}
where we defined new operators as
\begin{equation}
      a_f = a \sqrt{\frac{H(N)}{N}}=\sqrt{\frac{H(N+I)}{N+I}} a,\,\,\,\,\, a_f^+ = a^+\sqrt{\frac{H(N+I)}{N+I}} = \sqrt{\frac{H(N)}{N}} a^+ \label{ftrans}
      \end{equation}
      so that
      \begin{equation}
      a^+_f a_f = H(N), \,\,\,\,a_f a^+_f = H(N+I)
      \end{equation}
      and
      \begin{equation}
      [a_f, a_f^+] = H(N+I) - H(N).
      \end{equation}
      It gives the Hamiltonian
      \begin{equation}
      H = \frac{1}{2} [H(N) + H(N+I)]
      \end{equation}
 with discrete spectrum
 \begin{equation}
 E_n = \frac{1}{2} [H(n) + H(n+1)].
 \end{equation}
\subsection{Semi-relativistic oscillator as quantum f-oscillator}
As an example here we consider the semi-relativistic harmonic oscillator with energy
\begin{equation}
E(p) = \sqrt{m^2 c^4 + c^2 (p^2 + m^2 \omega^2 x^2)},
\end{equation}
in non-relativistic limit $c \rightarrow \infty$ becoming the usual harmonic oscillator
\begin{equation}
E(p) \approx m c^2 + \frac{1}{2 m}(p^2 + m^2 \omega_0^2 x^2) + O\left(\frac{1}{c}\right).
\end{equation}
Such type of dispersion appears in several physical applications. First is the problem of a relativistic electron in a magnetic field: the relativistic Landau
levels problem \cite{Rabi}. Another class of applications is the relativistic harmonic oscillator model of hadrons to explain magnetic moments of baryons and
hadron spectroscopy \cite{Bannur}.

In terms of action variables the Hamiltonian function is
\begin{equation}
H(J) = mc^2 \sqrt{1 + \frac{2\omega_0 }{m c^2 }\,J}
\end{equation}
and in terms of complex variables (\ref{complexcoordinate}) we have nonlinear oscillator
\begin{equation}
H(\alpha\bar\alpha) = mc^2 \sqrt{1 + \frac{2\omega_0 }{m c^2 }\, \alpha \bar\alpha},
\end{equation}
with nonlinear frequency
\begin{equation}
\omega(J) = \frac{\partial H(J)}{\partial J}= \frac{\omega_0}{\sqrt{1 + \frac{2\omega_0 }{m c^2 }\,J}}.
\end{equation}
In the non-relativistic limit it reduces to the linear oscillator frequency $\omega_0$.
Introducing
\begin{equation}
      \alpha_f = \sqrt{\frac{\sqrt{m^2c^4 + 2 m c^2 \omega_0 \alpha \bar\alpha}}{\alpha\bar\alpha}} \alpha,\,\,\,\,\,\bar \alpha_f = \sqrt{\frac{\sqrt{m^2c^4 + 2 m c^2 \omega_0 \alpha \bar\alpha}}{\alpha\bar\alpha}} \bar\alpha, \label{rotr}
      \end{equation}
with Poisson brackets
\begin{equation}
\{ \alpha_f, \bar \alpha_f  \} = - \rmi \omega_0\frac{mc^2}{|\alpha_f|^2},
\end{equation}
we get the classical f-oscillator form of the model
\begin{equation}
H(\alpha_f, \bar\alpha_f) = \alpha_f \bar\alpha_f.
\end{equation}
The evolution is determined by equation
\begin{equation}
\dot\alpha_f = - \rmi \omega(|\alpha_f|^2) \alpha_f,
\end{equation}
where the frequency is the nonlinear function depending on amplitude
\begin{equation}
\omega(|\alpha_f|^2) = \omega_0 \frac{mc^2}{|\alpha_f|^2}.
\end{equation}
When $c \rightarrow \infty$ it gives
\begin{equation}
\omega \approx \omega_0 (1 - \frac{\omega_0}{mc^2} |\alpha|^2) \label{dispersion1}
\end{equation}
so that the linear oscillator is recovered at $c = \infty$. The next order correction in $1/c$ is just a quadratic nonlinearity in amplitude.
Such type of nonlinearity is very specific and is described by cubic Nonlinear Schrodinger equation (NLS) as generic integrable nonlinear
envelope soliton equation \cite{Novikov}. It means that the first relativistic correction to the nonlinear dispersion (\ref{dispersion1}) will produce
NLS equation.

\subsubsection{Quantization of semi-relativistic oscillator}

Introducing operators
\begin{eqnarray}
      a_f = a\, \sqrt{\frac{\sqrt{m^2c^4 + 2 m c^2 \omega_0 N}}{N}}  = \sqrt{\frac{\sqrt{m^2c^4 + 2 m c^2 \omega_0 (N+1)}}{N+1}} \,a , \\
      a^+_f = a^+ \,\sqrt{\frac{\sqrt{m^2c^4 + 2 m c^2 \omega_0 (N+1)}}{N+1}} = \sqrt{\frac{\sqrt{m^2c^4 + 2 m c^2 \omega_0 N}}{N}}\, a^+ \label{routr}
      \end{eqnarray}
we find relations
\begin{equation}
a^+_f a_f = \sqrt{m^2c^4 + 2 m c^2 \omega_0 N},\,\,\,\,a_f a^+_f = \sqrt{m^2c^4 + 2 m c^2 \omega_0 (N+1)}
\end{equation}
and commutator
\begin{equation}
[a, a^+] = \sqrt{m^2c^4 + 2 m c^2 \omega_0 (N+1)} - \sqrt{m^2c^4 + 2 m c^2 \omega_0 N}.
\end{equation}
Then the  Hamiltonian is
\begin{equation}
H = \frac{1}{2}[\sqrt{m^2c^4 + 2 m c^2 \omega_0 (N+1)} - \sqrt{m^2c^4 + 2 m c^2 \omega_0 N}],
\end{equation}
with discrete spectrum
\begin{equation}
E_n = \frac{1}{2}[\sqrt{m^2c^4 + 2 m c^2 \omega_0 (n+1)} - \sqrt{m^2c^4 + 2 m c^2 \omega_0 n}].
\end{equation}

\section{Classical and quantum symmetric q-oscillator}
Here we briefly reproduce main formulas for the classical and quantum $q$-oscillator \cite{Manko1}, \cite{Manko2}.
\subsection{Classical symmetric q-oscillator} In terms of complex variables 
\begin{equation}
\alpha = \frac{x + \rmi p}{\sqrt{2}},\,\,\,\bar\alpha = \frac{x - \rmi p}{\sqrt{2}}
\end{equation}
the Hamilton function is
\begin{equation}
H_q (\alpha, \bar\alpha) = \frac{\sinh (\lambda \bar \alpha \alpha)}{\sinh \lambda},
\end{equation}
with Poisson bracket
\begin{equation}
\{ \alpha, \bar \alpha\} = -\rmi,
\end{equation}
and evolution equations
\begin{equation}
\dot\alpha = - \rmi \omega_q \alpha,\,\,\,\dot{\bar\alpha} =  \rmi \omega_q \bar\alpha.
\end{equation}

The action-angle variables are defined as
\begin{equation}
\alpha = \rmi \sqrt{J} \rme^{-\rmi \theta},\,\,\,\,\bar \alpha = -\rmi \sqrt{J} \rme^{\rmi \theta}
\end{equation}
with bracket $\{ \theta, J \} = 1$, so that the Hamiltonian function depends only on $J$,
\begin{equation}
H_q (J) = \frac{\sinh (\lambda J)}{\sinh \lambda}.
\end{equation}
Then the nonlinear frequency depends exponentially on amplitude and the energy
\begin{equation}
\omega_q(J) = \frac{\partial H(J)}{\partial J}= \frac{\lambda}{\sinh \lambda} \cosh (\lambda J).
\end{equation}

\subsubsection{Quantum symmetric q oscillator}
The standard creation and annihilation operators $a$ and $a^+$ with commutator $[a, a^+] = 1$ in the Fock basis $\{ |n>\}$, $n = 0,1,2...$
determines the number operator $N = a^+ a$ and symmetric q-number operator
\begin{equation}
[N]_{\tilde q} = \frac{q^N - q^{-N}}{q - q^{-1}} = \frac{\sinh \lambda N}{\sinh \lambda},
\end{equation}
 where $q \equiv \rme^{\lambda}$. By transformation \cite{Song}, \cite{Kulish}, \cite{Manko1}
\begin{equation}
      a_f = a \sqrt{\frac{[N]_{\tilde q}}{N}}=\sqrt{\frac{[N+I]_{\tilde q}}{N+I}} a,\,\,\,\,\, a_f^+ = a^+\sqrt{\frac{[N+I]_{\tilde q}}{N+I}} = \sqrt{\frac{[N]_{\tilde q}}{N}} a^+ \label{ftransformation}
      \end{equation}
      so that
      \begin{equation}
      a^+_f a_f = [N]_{\tilde q}, \,\,\,\,a_f a^+_f = [N+I]_{\tilde q}
      \end{equation}
      and
      \begin{equation}
      [a_f, a_f^+] = [N+I]_{\tilde q} - [N]_{\tilde q}
      \end{equation}
the Hamiltonian is represented in terms of symmetric $q$-number operators
\begin{equation}
      H = \frac{a_f a_f^+ + a_f^+ a_f}{2}=\frac{1}{2} ([N]_{\tilde q} + [N+I]_{\tilde q})
      \end{equation}
and the  spectrum is \cite{Biedenharn}, \cite{Macfarlane}, \cite{Sun}
\begin{equation}
E_n =\frac{1}{2} ([n]_{\tilde q} + [n+1]_{\tilde q}) =\frac{1}{2} \frac{\sinh ((n+\frac{1}{2})\lambda)}{\sinh \frac{\lambda}{2}}.
\end{equation}

\section{The Schr\"odinger and complex Burgers equations for symmetric q-oscillator dispersion}

\subsection{q-Schr\"odinger equation}
The first quantization of the $q$-oscillator is described by the linear Schr\"odinger equation
 with Hamiltonian
\begin{equation}
\hat H_q = \frac{1}{\sinh \lambda} \sinh \frac{\lambda}{2}\left(- \frac{\hbar^2}{m} \frac{\rmd^2}{\rmd x^2} + k x^2 \right).\label{qoscillatorSchrodinger}
\end{equation}

For free q-particle, when the oscillator coupling constant $k = 0$, we have the Hamiltonian
\begin{equation}
\hat {H_0}_q = \frac{1}{\sinh \lambda} \sinh \left(- \frac{\lambda \hbar^2}{2 m} \frac{\rmd^2}{\rmd x^2} \right),\label{qhamiltonian}
\end{equation}
which determines the linear Schrodinger equation with nonlinear dispersion

\begin{equation}
\rmi \hbar \frac{\partial \psi}{\partial t} = \frac{1}{\sinh \lambda} \sinh \left(- \frac{\lambda \hbar^2}{2 m} \frac{\partial^2}{\partial x^2} \right) \psi.
\label{linearqSchrodinger}\end{equation}
We can call it as a q-Schrodinger equation.
\subsection{Symmetry operators and solutions}
Operators of the time and space translations
\begin{equation}
P_0 = \rmi\hbar \frac{\partial}{\partial t},\,\,\,  P_1 = -\rmi\hbar \frac{\partial}{\partial x},
\end{equation}
commute with
the Schr\"odinger operator
\begin{equation}
S = \rmi \hbar \frac{\partial }{\partial t} - \hat {H_0}_q  = \rmi \hbar \frac{\partial }{\partial t} - \frac{1}{\sinh \lambda} \sinh \left(- \frac{\lambda \hbar^2}{2 m} \frac{\partial^2}{\partial x^2} \right),
\end{equation}
 $[P_\mu, S] = 0$, $\mu = 0,1$.
The boost operator
\begin{equation}
K = x + \frac{\rmi\hbar t}{m} \frac{\lambda}{\sinh \lambda} \cosh \frac{\lambda}{2}\left(- \frac{\hbar^2}{m} \frac{d^2}{dx^2} \right) \frac{d}{dx}\label{boost}
\end{equation}
is also commuting with $S$, $[K,S] = 0$.
Commuting it with space and time translations we obtain the following algebra
of symmetry operators
\begin{equation}
[P_0, P_1] = 0,\,\,[P_1,K] = - \rmi\hbar,\,\,[P_0, K] = -\frac{ \lambda}{m \sinh \lambda} \cosh \frac{\lambda}{2}\left( \frac{P_1}{m} \right) P_1.\label{symmetryalgebra}
\end{equation}
An operator commutative with Schrodinger operator determines the dynamical symmetry \cite{Malkin}.
If $\Psi$ is a solution of (\ref{Schrodinger}) and $W$ is an operator, so that $[W, S] = 0$  then $S \Psi$ is
also a solution of  (\ref{Schrodinger}). Then as follows $K$ operator can generate new solutions of (\ref{Schrodinger}).

In the limit $\lambda \rightarrow 0$ the boost operator (\ref{boost}) reduces to the Galilean boost
\begin{equation}
K = x + \frac{\rmi\hbar t}{m} \frac{\rmd}{\rmd x}
\end{equation}
and $q$-deformed symmetry algebra (\ref{symmetryalgebra}) to the usual non-relativistic algebra of the Galilean group \cite{Malkin}.

\subsubsection{q-Polynomial solutions}

For given classical dispersion
\begin{equation}E(p) = \frac{1}{\sinh \lambda} \sinh \left(\lambda \frac{p^2}{2 m} \right)
\end{equation}
 we define the plane wave solution as a generating function of $q$-Kampe de Feriet polynomials
  $H^{(q)}_n (x,t)$, $q = \rme^{\lambda}$,
\begin{equation}
\rme^{\frac{\rmi}{\hbar}( p x - E(p)  t)} = \sum^\infty_{n=0} \left(\frac{\rmi}{\hbar}\right)^n \frac{p^n}{n!} H^{(q)}_n(x,t).
\end{equation}
The polynomials
\begin{equation}
H^{(q)}_n(x,t) = \exp {\left(-\frac{\rmi t}{\hbar}\frac{1}{\sinh \lambda} \sinh \left(- \frac{\lambda \hbar^2}{2 m} \frac{\rmd^2}{\rmd x^2} \right) \right)}\, x^n \label{qKF}
\end{equation}
are solutions of (\ref{Schrodinger}) with initial value $H^{(q)}_n(x,0) = x^n$.
From commutativity $[S,K] = 0$, operator $K$ evolves according to
the Heisenberg equation
\begin{equation}
\rmi\hbar \frac{\partial K}{\partial t} = [\hat {H_0}_q , K]
\end{equation}
and has the form
\begin{equation}
K(t) = \rme^{-\frac{\rmi}{\hbar}\hat {H_0}_q  t}K(0)\, \rme^{\frac{\rmi}{\hbar}\hat {H_0}_q  t} = \rme^{-\frac{\rmi}{\hbar}\hat {H_0}_q  t} x \,\rme^{\frac{\rmi}{\hbar}\hat {H_0}_q  t}.
\end{equation}
From this follows that operator $K$ generates an infinite hierarchy of polynomial solutions according to the recursion
\begin{equation}
K H^{(q)}_n(x,t) = K \rme^{-\frac{\rmi}{\hbar}\hat {H_0}_q  t}x^n =  \rme^{-\frac{\rmi}{\hbar}\hat {H_0}_q  t}x^{n+1} = H^{(q)}_{n+1}(x,t),
\end{equation}
and first few polynomials are
\begin{eqnarray}
H^{(q)}_1 = x,\\
H^{(q)}_2 = x^2 + \rmi \frac{\hbar}{m}\frac{\lambda}{\sinh \lambda} \,t, \\
H^{(q)}_3 = x^3 + 3\,\rmi\frac{\hbar}{m} \frac{\lambda}{\sinh \lambda}\, x\, t, \\
H^{(q)}_4 = x^4 + 6\,\rmi\frac{\hbar}{m} \frac{\lambda}{\sinh \lambda}\, x^2\, t - 3 \frac{\hbar^2}{m^2} \frac{\lambda^2}{\sinh^2 \lambda}\,t^2, \\
H^{(q)}_5 = x^5 + 10\,\rmi\frac{\hbar}{m} \frac{\lambda}{\sinh \lambda}\, x^3\, t - 15 \frac{\hbar^2}{m^2} \frac{\lambda^2}{\sinh^2 \lambda}\,x \,t^2,\\
H^{(q)}_6 = x^6 + 15\,\rmi\left(\frac{\hbar}{m} \frac{\lambda}{\sinh \lambda}\right)\, x^4\, t - 45 \left(\frac{\hbar}{m} \frac{\lambda}{\sinh \lambda}\right)^2\,x^2 \,t^2 \\
-15 \,\rmi \left(\frac{\hbar}{m} \frac{\lambda}{\sinh \lambda}\right)^3 t^3 + 30 \rmi \left(\frac{\hbar}{m} \frac{\lambda}{\sinh \lambda}\right) \frac{\lambda^2 \hbar^4}{m^2} t.
\end{eqnarray}
In the limit $q \rightarrow 1$ or $\lambda \rightarrow 0$ we have  $\lambda/\sinh \lambda \rightarrow 1$, and the above polynomials reduce to the Schrodinger polynomials \cite{Pashaev3}
 \begin{equation}
H^{(S)}_n(x,t) = \rme^{\frac{\rmi}{\hbar}t \frac{\hbar^2}{2m} \frac{\rmd^2}{\rmd x^2}} x^n.
\end{equation}
Let $H^{(KF)}_n(x,t) = \exp [t \frac{\rmd^2}{\rmd x^2}] x^n$ are the standard Kampe de Feriet polynomials, then
$H^{(S)}_n (x,t) = H^{KF} (x, \frac{\rmi\hbar}{2m}t)$
or in terms of the Hermit polynomials
\begin{equation}
H^{(S)}(x,t) = \left( -\frac{\rmi\hbar}{2m}t\right)^{n/2} H_n \left(\frac{x}{\sqrt{-2\rmi\hbar t/m}}\right).
\end{equation}
We notice that for $n=1,...,5$ only coefficients in $H^{(q)}_n$ becomes deformed, while starting from $n=6$ a new term, vanishing in the limit $\lambda \rightarrow 0$ appears.

 \subsubsection{Motion of zeros}

Motion of N zeros for polynomial solutions (\ref{qKF}) is determined by the system of ordinary differential equations
\begin{equation}
\dot x_k = \frac{\rmi}{\hbar\, \sinh \lambda} {\rm Res}|_{x= x_k} \sinh \left[\frac{\lambda}{2 m}\frac{\hbar^2}{i^2}\left( \frac{\rmd}{\rmd x} + \sum^N_{l=1}
 \frac{1}{x - x_l}\right)^2 \right] \cdot 1,
\end{equation}
k=1,...,N, describing system of N interacting particles in a line.
For $\lambda \rightarrow 0$ the polynomials of complex argument, correspond to holomorphic extension of Schrodinger equation in (2+1)-dimensional Chern-Simons theory \cite{PG} and zeros of these polynomials describe motion
of point vortices in plane. Then for $\lambda \neq 0$, consideration of holomorphic q-Schrodinger equation could describe a q-deformed vortex dynamics in the plane.
\section{Complex q-Burgers equation}
Using Schr\"odinger's log $\Psi$ transform : $\Psi = \rme^{\ln \Psi}$ and identity
 \begin{equation}
 \rme^{- \ln \Psi} \frac{\partial^n}{\partial x^n} \rme^{\ln \Psi} = \left(\frac{\partial}{\partial x} + \frac{\partial \ln \Psi}{\partial x} \right)^n
\end{equation}
the q-Schr\"odinger equation (\ref{Schrodinger}) can be rewritten in the form
\begin{equation}
\rmi\hbar \frac{\partial }{\partial t} \ln \Psi = \frac{1}{\sinh \lambda} \sinh \left( -\lambda  \frac{\hbar^2}{2 m} \left( \frac{\partial}{\partial x} + \frac{\partial \ln \Psi}{\partial x}\right)^2 \right)\cdot 1 .\label{schrodinger3}
\end{equation}
For a given complex function $\Psi = \rme^{\frac{\rmi}{\hbar}F} = \rme^{R + \frac{\rmi}{\hbar} S}$ we
introduce a new complex function with dimension of velocity
\begin{equation}
V = - \rmi \frac{\hbar}{m} \frac{\partial}{\partial x} \ln \Psi = \frac{1}{m}F,
\end{equation}
where $F = S - \rmi\hbar R$ is the complex potential, with real and imaginary parts as the  classical and quantum velocities
\begin{equation}
V = V_c + \rmi V_q = \frac{1}{m} S_x - \rmi\frac{\hbar}{m}R_x.
\end{equation}

Then (\ref{schrodinger3}) becomes the quantum q-Hamilton-Jacobi equation:
\begin{equation}
\frac{\partial F}{\partial t} + \frac{1}{\sinh \lambda} \sinh \left(\frac{\lambda}{2 m}\left(-\rmi\hbar \frac{\partial}{\partial x} + F_x\right)^2
\right)\cdot 1 = 0. \label{CHJ}
\end{equation}
In the classical (dispersionless) limit $\hbar \rightarrow 0$ the quantum velocity $V_q$ vanishes and the complex potential reduces to the
real velocity potential $F \rightarrow S$, so that (\ref{CHJ}) becomes the classical q-Hamilton-Jacobi equation for action $S$:
\begin{equation}
\frac{\partial S}{\partial t} +  \frac{1}{\sinh \lambda} \sinh \left(\lambda\frac{1}{2 m}\left(\frac{\partial S}{\partial x}\right)^2
\right)= 0. \label{ClHJ}
\end{equation}
Differentiating (\ref{schrodinger3}) we have nonlinear complex q-Burgers type equation for the complex velocity (complex q-Madelung equation)
\begin{equation}
\frac{\partial V}{\partial t} + \frac{1}{m \sinh \lambda} \frac{\partial}{\partial x}\sinh \left(\frac{\lambda}{2 m}\left(-\rmi\hbar \frac{\partial}{\partial x} + m V\right)^2
\right)\cdot 1 .\label{schrodinger4}
\end{equation}
In the classical limit it gives the Newton equation
 in the hydrodynamic form
\begin{equation}
\frac{\partial V_c}{\partial t} + \frac{\lambda}{\sinh \lambda} V_c \cosh \left( \lambda \frac{m V^2_c}{2}\right) \frac{\partial V_c}{\partial x} =0, \label{HF}
\end{equation}
which is just differentiation of the classical Hamilton-Jacobi equation (\ref{ClHJ}). Equation (\ref{HF})
has implicit general solution
\begin{equation}
V_c(x,t) = f\left(x -   V_c\, t  \frac{\lambda}{\sinh \lambda} \cosh \left( \lambda \frac{m V^2_c}{2}\right) \right)
\end{equation}
where $f$ is an arbitrary function, and develops shock at a critical time when derivative
$(V_c)_x$
is blowing up.
\subsection{B\"acklund transformation for complex q-Burgers equation}

As we have seen, using boost transformation (\ref{boost}) from a given solution $\Psi_1$ of the q-Schr\"odinger equation
(\ref{Schrodinger}) we can generate another solution as
\begin{equation}
\Psi_2 = K \Psi_1 = \left[ x + \frac{\rmi\hbar t}{m} \frac{\lambda}{\sinh \lambda} \cosh \left(\lambda\frac{\hbar^2}{2 m} \frac{\rmd^2}{\rmd x^2} \right) \frac{\rmd}{\rmd x}\right] \Psi_1.
\end{equation}
By using identity
\begin{equation}
\Psi^{-1} G\left(-\rmi\hbar \frac{\partial}{\partial x}\right) \Psi = G\left(-\rmi\hbar \frac{\partial}{\partial x} + m V\right)\cdot 1\end{equation}
for complex velocities $V_a = -\rmi \frac{\hbar}{m} \ln \Psi_a$, $(a=1,2)$, for complex q-Burger equation (\ref{schrodinger4}) we obtain the B\"acklund transformation
\begin{equation}
V_2 = V_1 - \rmi \frac{\hbar}{m} \frac{\partial}{\partial x} \ln \left[x - t \frac{\lambda}{\sinh \lambda}\cosh \left(\frac{\lambda}{2 m}\left(-\rmi\hbar \frac{\partial}{\partial x} + m V \right)^2\right)\cdot V_1\right].
\end{equation}

It is noted that in the classical limit $\hbar \rightarrow 0$, $V \rightarrow V_c$ the above B\"acklund transformation  reduces to the trivial identity ${V_c}_1 = {V_c}_2$.

 \section{Integrable nonlinear q-Schr\"odinger equation}

In previous sections we have studied q-Schr\"odinger equation determined by hyperbolic sine dispersion.
By adding the oscillator coupling term it becomes exactly solvable q-oscillator equation (\ref{qoscillatorSchrodinger}).
 Here we show that from this dispersion it is
 possible to construct also NLS type integrable evolution equation .
This q-NLS equation in the limit $q \rightarrow 1$ or $\lambda \rightarrow 0$ reduces to the NLS equation. It is interesting to note that by expansion in powers of $\lambda$ then
we have an infinite set of NLS type equations integrable at arbitrary order of deformation $q$.
\section{q-nonlinear Schr\"odinger equation}

 Dispersion formula for linear q-Schr\"odinger equation (\ref{linearqSchrodinger})  expanded to series in $p^2$ is
 \begin{equation}
 E(p) = \frac{1}{\sinh \lambda} \sinh \left(\lambda \frac{p^2}{2m}\right) = \frac{\lambda}{\sinh \lambda} \frac{p^2}{2m} + \frac{\lambda^3}{3! \sinh \lambda} \left(\frac{p^2}{2m}\right)^3 + ...
\end{equation}
 This may be used to construct a linear q-Schr\"odinger
equation as a formal power series
\begin{eqnarray}
 \rmi\hbar \frac{\partial}{\partial t}\psi =\left[\frac{\lambda}{\sinh \lambda} \left(\frac{-\hbar^2}{2m}\frac{\partial^2}{\partial x^2}\right) + \frac{\lambda^3}{3! \sinh \lambda} \left(\frac{-\hbar^2}{2m}\frac{\partial^2}{\partial x^2}\right)^3 + ...\right]
  \psi
 \end{eqnarray}
 Combining two complex conjugate versions of this equation together we have the system
\begin{equation} \rmi\sigma_3  \left (\begin{array}{clcr}\psi \\
\bar\psi \end{array} \right)_{t}= \frac{1}{\sinh \lambda} \sinh \left(- \frac{\lambda \hbar^2}{2 m}  \left(\rmi\sigma_3 \frac{\partial}{\partial x}\right)^{2}  \right) \left(\begin{array}{clcr} \psi\\
\bar\psi \end{array} \right). \label{RelLS}\end{equation}

Following the general procedure described in Appendix A one may proceed further: by replacing the derivative
operator ${\cal R}_0 = \rmi\sigma_3\frac{\partial}{\partial x}$ as a 
momenta to the full recursion operator $\cal{R}$
(\ref{recursion}), one obtains an {\it integrable}
nonlinear q-Schrodinger equation (q-NLS)
\begin{equation} \rmi\sigma_3  \left (\begin{array}{clcr}\psi \\
\bar\psi \end{array} \right)_{t}= \frac{1}{\sinh \lambda} \sinh \left(- \frac{\lambda \hbar^2}{2 m}  {\cal{R}}^{2}  \right) \left(\begin{array}{clcr} \psi\\
\bar\psi \end{array} \right) . \label{qNLS}\end{equation}

\subsubsection{The linear problem}

Applying general result of Appendix A to the above q-deformed non-relativistic
dispersion, we have the next linear problem (the Lax representation) for equation
(\ref{qNLS})
\begin{equation}
\frac{\partial }{\partial x}\left(\begin{array}{c} v_1\\
v_2\end{array} \right) = \left( \begin{array}{cc} -
\frac{\rmi}{2}p& -\kappa^2 \bar\psi \\ \psi &\frac{\rmi}{2}p
\end{array}\right)\left(\begin{array}{c} v_1\\
v_2\end{array} \right) , \label{ZS1RelNLS}\end{equation}

\begin{equation}
\frac{\partial }{\partial t}\left(\begin{array}{c} v_1\\
v_2\end{array} \right) =  \left(
\begin{array}{cc} - \rmi A& -\kappa^2 \bar C \\ C & -\rmi A
\end{array}\right)\left(\begin{array}{c} v_1\\
v_2\end{array} \right),\label{ZS2RelNLS}
\end{equation}

where
\begin{equation} \left( \begin{array}{c} C\\ \bar C
\end{array}\right)=
  \frac{\sinh \frac{\lambda}{2m}{\cal{R}}^2 - \sinh \frac{\lambda}{2m}{p}^2}{({\cal{R}} - p) \sinh \lambda}
  \left(\begin{array}{c} \psi\\ \bar \psi
\end{array}\right), \label{CRel}
\end{equation}
\begin{equation}
\fl A =   -\frac{1}{2}\sinh \frac{\lambda}{2m}{p}^2 - \rmi\kappa^2 \left(
\int^x \bar\psi, - \int^x \psi \right) \frac{\sinh \frac{\lambda}{2m}{\cal{R}}^2 - \sinh \frac{\lambda}{2m}{p}^2}{({\cal{R}} - p) \sinh \lambda}\left(
\begin{array}{c} \psi\\ \bar \psi
\end{array}\right)\label{ARel}
\end{equation}
and the spectral parameter $p$ has meaning of the classical
momentum.
The model 
 is an integrable nonlinear
q-Schrodinger equation with q-deformed dispersion as well as the nonlinear terms.
In the limit $\lambda \rightarrow 0$ it reduces to standard NLS model (\ref{NLS}). The last one as a universal
integrable equation describing envelope modulation has appeared in many applications, especially in
nonlinear optics. It admits N-soliton solutions, higher symmetries etc.\cite{Novikov}
Remarkable property of our model (\ref{qNLS}) is that it generalizes the NLS model in a very special way. If we expand it in
$\lambda$, then at every order of $\lambda$  we get an integrable
system. It means that  we have integrable corrections
to the NLS equation at any order of $\lambda$.
Due to the symmetry $\lambda \rightarrow -\lambda$ of (\ref{qNLS}) only even powers of $\lambda$ appear in the expansion.
Then the lowest corrections are given by integrable NLS model with six order dispersion $\psi_{xxxxxx}$ and up to 13-th order nonlinearity
 \begin{equation}\fl \rmi\sigma_3  \left (\begin{array}{clcr}\psi \\
\bar\psi \end{array} \right)_{t}= \frac{\hbar^2}{2 m}  {\cal{R}}^{2}   \left(\begin{array}{clcr} \psi\\
\bar\psi \end{array} \right)  +  \frac{\lambda^2}{6}\left[-\frac{\hbar^2}{2 m}  {\cal{R}}^{2} + \frac{\hbar^6}{(2 m)^3}  {\cal{R}}^{6}\right]  \left(\begin{array}{clcr} \psi\\
\bar\psi \end{array} \right) + O(\lambda^4).
\label{qcorrectionNLS}\end{equation}

 \section{Hydrodynamic flow in bounded domain}
As a next variation on the subject here we consider hydrodynamical problem related with motion in double connected domain.
This problem can naturally be formulated in terms of q-calculus  and as we will see, for the point vortex motion in annulus it provides an example of nonlinear oscillator, formulated as an $f$-oscillator.
\subsection{ Hydrodynamic flow in bounded domain}
For incompressible and irrotational flow in a domain bounded by curve $C$, the problem is to
find analytic function $F(z)$ with
boundary condition
$$ \Im F|_C =  \psi|_{C} = 0, $$
where $\psi$ is the stream function. Then the normal velocity vanishes at boundary
$v_{n}|_{C} = 0$.

\subsection{ Milne-Thomson's one circle theorem}
For a given flow in plane with complex potential $f(z)$,
introduction of boundary circle at the origin $C$: $|z| = r$
produces the flow with
complex potential  \cite{MilneThomson}
$$F(z) = f(z) + \bar f \left(\frac{r^2}{z}\right). $$

\subsection{Two circles theorem}

For annular domain, $r_1 <|z|< r_2$, between two concentric circles $C_1: |z| = r_1$, $C_2: |z| = r_2$
the complex potential is \cite{Pashaev2}
\begin{equation}F_{Q}(z) = f_{Q}(z) + \bar f_{Q} \left(\frac{r^2}{z}\right),\label{twocircle}\end{equation}
where ${Q = \frac{r^2_2}{r^2_1}}$,
$ f_{Q}(z) = \sum^\infty_{n=-\infty} f({Q}^n z)$ is flow in even annular image domain,
$\bar f_{Q} \left(\frac{r^2}{z}\right) = \sum^\infty_{n=-\infty} \bar f \left({ Q}^n \frac{r^2}{z}\right)$ is the flow in odd annular image domain.

From (\ref{twocircle})  follows that
$f_{Q}({Q} z) = f_{Q}(z)$, which implies that the
complex potential is Q-periodic function $F_{Q}({Q} z) = F_{Q}(z)$.
Depending on number and position of vortices or other objects, we fix singularity of this function in terms of q-elementary functions \cite{PY1},\cite{Pashaev2}.
\subsection{Wedge theorem}
Here we are going to formulate some new theorems in wedge domain.
For a given flow in plane with complex potential $f(z)$,
introduction of boundary wedge with angle $\alpha = 2\pi/N = \pi/n$, where $N = 2n$ is a positive even number,
 produces the flow with
complex potential
\begin{eqnarray}F_q(z)& =& f(z) + f(q^2 z)  + f(q^4 z) + ...+ f(q^{2(n-1)}z) \\ &+& \bar f(z) + \bar f(q^2 z) + \bar f(q^4 z)+ ...+ \bar f(q^{2(n-1)} z) \label{wedge1}\end{eqnarray}
or shortly
\begin{equation}F_q(z) = \sum^{n-1}_{k=0} f(q^{2 k}z) + \sum^{n-1}_{k=0} \bar f(q^{2 k}z), \label{wedge2}\end{equation}
where $q = \rme^{\rmi\frac{2 \pi}{N}} = \rme^{\rmi\frac{ \pi}{n}}$ is the primitive root of unity: $q^N = q^{2n}= 1$.
For proof of this theorem see Appendix B.

Here we notice that the complex potential (\ref{wedge2}) is $q^2$ periodic analytic function
$F_q(q^2 z) = F_q(z)$, while the complex velocity $\bar V(z) = dF_q(z)/dz$ is the scale invariant analytic function
$\bar V(q^2 z) = q^{-2} \bar V(z)$.

\subsubsection{Vortex kaleidoscope}
As an example, we consider single vortex in the wedge at point $z_0$:
\begin{equation}
f(z) = \frac{\rmi \Gamma }{2 \pi} \ln (z - z_0).
\end{equation}
Then applying the above wedge theorem we get complex potential
\begin{equation}
F_q(z) = \frac{\rmi \Gamma }{2 \pi} \sum^{n-1}_{k=0}\ln \frac{z- z_0 q^{2k}}{z - \bar z_0 q^{2k}} = \frac{\rmi \Gamma }{2 \pi} \ln \prod^{n-1}_{k=0}\frac{z- z_0 q^{2k}}{z - \bar z_0 q^{2k}},
\end{equation}
describing a kaleidoscope of $2n$ vortices with positive strength at points $z_0, z_0 q^2, z_0 q^4...,$ $ z_0 q^{2(n-1)} $ and with negative
strength at points $\bar z_0, \bar z_0 q^2, \bar z_0 q^4...,$  $ \bar z_0 q^{2(n-1)} $. Positive images are just rotations of original vortex
position $z_0$ on angles $\frac{2\pi}{n}, 2 \frac{2\pi}{n}, 4 \frac{2\pi}{n}, ..., (n-1)\frac{2\pi}{n} $, while negative images are rotations of the
reflected vortex position $\bar z_0$ on the same angles. This expression can be drastically simplified due to the next identity
\begin{equation}
(z - z_0) (z - z_0 q^{2})(z - z_0 q^{4})...(z- z_0 q^{2(n-1)}) = z^n - z^n_0.
\end{equation}
This identity is valid for $q^2 = \rme^{\rmi \frac{2\pi}{n}}$ as the primitive $n$-th root of unity and has been considered long time ago by E. Kummer.
 Simplest proof follows from factorization of polynomial $z^n - z^n_0$ by roots of unity. As a result, finally we get the following compact expression for the
 vortex flow in the wedge (we can call it as the Kummer kaleidoscope of vortices)
 \begin{equation}
 F_q(z) = \frac{\rmi \Gamma }{2 \pi} \ln \frac{z^n- z_0^n}{z^n - \bar z_0^n}.\label{kummer}
 \end{equation}

 For $n=1$ we have vortex at $z_0$ in upper half plane, with one image at $\bar z_0$
 \begin{equation}
 F(z) = \frac{\rmi \Gamma }{2 \pi} \ln \frac{z- z_0}{z - \bar z_0}.
 \end{equation}
 For $n=2$ the vortex at $z_0$ in first quadrant produces images at $-z_0, \bar z_0, -\bar z_0$
 \begin{equation}
 F(z) = \frac{\rmi \Gamma }{2 \pi} \ln \frac{z^2- z_0^2}{z^2 - \bar z_0^2} =\frac{\rmi \Gamma }{2 \pi} \ln \frac{(z- z_0)(z+ z_0)}{(z - \bar z_0)(z + \bar z_0)} .
 \end{equation}

\subsection{Circular wedge theorem}
Here we consider circular wedge with angle $\alpha = 2\pi/N = \pi/n$, bounded by lines $\Gamma_1$: $z = x $ and $\Gamma_2$: $z = x \rme^{\rmi\frac{\pi}{n}}$
and the circular boundary $C_1:$ $z = r \rme^{\rmi t}$, $0 < t < \alpha$. Then, the  flow bounded by such domain is
\begin{equation}
F_q(z) = \sum^{n-1}_{k=0}[ f(q^{2 k}z) + \bar f(q^{2 k}z) ]+ \sum^{n-1}_{k=0}[ \bar f(\frac{r^2}{q^{2 k}z}) +  f(\frac{r^2}{q^{2 k}z})].
\end{equation}
The proof is similar to the above one and shows that imaginary part of $F(z)$ vanishes at boundaries $\Gamma_1$, $\Gamma_2$ and $C_1$. The theorem could be considered as combination of Milne-Thomson's one circle theorem with the wedge theorem.

\subsubsection{Doubled vortex kaleidoscope}
For single vortex in circular wedge after some calculations and simplifications we obtain
\begin{equation}
 F_q(z) = \frac{\rmi\Gamma}{2 \pi} \ln \frac{(z^n- z_0^n)(z^n - \frac{r^{2n}}{z^n_0})}{(z^n - \bar z_0^n)(z^n - \frac{r^{2n}}{\bar z^n_0})}.
 \end{equation}
Comparing with vortex kaleidoscope (\ref{kummer}) we observe doubling of images by reflection in circle $r$.
\subsection{Double circular wedge theorem}
Now we consider
more general region, the double circular wedge, bounded by two lines $\Gamma_1$: $z = x $ and $\Gamma_2$: $z = x \rme^{\rmi\frac{\pi}{n}}$
and two circular boundaries $C_1:$ $z = r_1 \rme^{\rmi t}$, $0 < t < \alpha$, and $C_2:$ $z = r_2 \rme^{\rmi t}$, $0 < t < \alpha$.
By combination of two circle theorem with the wedge theorem we have the flow
\begin{equation}
F(z) = f_{qQ}(z) + \bar f_{qQ}(z) + \bar f_{qQ} \left(\frac{r^2_2}{z}\right) + f_{qQ} \left(\frac{r^2_2}{z}\right),
\end{equation}
where
\begin{equation}
f_{qQ}(z) \equiv \sum^\infty_{m=-\infty}\sum^{n-1}_{k=0} f(Q^m q^{2k} z).
\end{equation}
In explicit form we obtain
\begin{equation}
F(z) = \sum^\infty_{m=-\infty}\sum^{n-1}_{k=0} [f(Q^m q^{2k}z) + \bar f(Q^m q^{2k} z) + \bar f\left(Q^m q^{2k}\frac{r^2_2}{z}\right) + f \left(Q^m q^{2k}\frac{r^2_2}{z}\right)]
\end{equation}
It is noticed that in this case we have q-calculus with two different basis. The first one $Q = r^2_2/r^2_1$ is a real number relating an infinite number of reflections in both circular boundaries. The second one $q^2 = \rme^{\rmi \frac{2\pi}{n}}$ is a complex unitary number with finite number of reflections $n$.
The complex potential here is a double $q,Q$ - periodic analytic function:
\begin{equation}
F(q^2 z) = F(z),\,\,\,\,\,\,F(Q z)= F(z).
\end{equation}

 \subsubsection{Self-similar infinite vortex kaleidoscope}
For single vortex in the double circular wedge we get result
\begin{equation}
 F(z) = \frac{\rmi \Gamma}{2 \pi} \sum^{\infty}_{m= -\infty}\ln \frac{(z^n- z_0^n Q^{nm})(z^n - \frac{r^{2n}_2}{z^n_0}Q^{nm})}{(z^n - \bar z_0^n Q^{nm})(z^n - \frac{r^{2n}_2}{\bar z^n_0} Q^{nm})}.
\end{equation}
This function
 describes self-similar kaleidoscope of infinite set of vortices on $Q$ geometric lattice. It generalizes expression for single vortex images in
 concentric annular domain considered in \cite{PY1}.
 For $n = 1$, $q^2 =1$, and we have single vortex in upper half-plane of annular domain
\begin{equation}
 F(z) = \frac{\rmi \Gamma}{2 \pi} \sum^{\infty}_{m= -\infty}\ln \frac{(z- z_0 Q^{m})(z - \frac{r^{2}_2}{z_0}Q^{m})}{(z - \bar z_0 Q^{m})(z - \frac{r^{2}_2}{\bar z_0} Q^{m})}.\label{annularvortex}
\end{equation}
For $n=2$ , $q^2 = -1$ we have single vortex in first quadrant of annulus and
\begin{equation}
 F(z) = \frac{\rmi \Gamma}{2 \pi} \sum^{\infty}_{m= -\infty}\ln \frac{(z^2- z_0^2 Q^{2m})(z^2 - \frac{r^{4}_2}{z^2_0}Q^{2m})}{(z^2 - \bar z_0^2 Q^{2m})(z^2 - \frac{r^{4}_2}{\bar z^2_0} Q^{2m})},
\end{equation}
or
\begin{equation}
 F(z) = \frac{\rmi \Gamma}{2 \pi} \sum^{\infty}_{m= -\infty}\ln \frac{(z- z_0 Q^{m})(z+ z_0 Q^{m})(z - \frac{r^{2}_2}{z_0}Q^{m})(z + \frac{r^{2}_2}{z_0}Q^{m})}{(z - \bar z_0 Q^{m})
 (z + \bar z_0 Q^{m})
 (z - \frac{r^{2}_2}{\bar z_0} Q^{m})(z + \frac{r^{2}_2}{\bar z_0} Q^{m})}.
\end{equation}
This formula demonstrates how vortices are reflected for every value of $m$.

\section{Vortex in annular domain as f-oscillator}
As an application of above formulas
 here we consider the point vortex problem in annular domain
as a nonlinear oscillator, or as the specific form of the f-oscillator.

 \subsection{Vortex rotation in annulus as nonlinear q-oscillator}

For a single
vortex in annular domain the complex velocity at the vortex position
is determined by \cite{PY1}, \begin{equation} \dot{z}_0 = \dot{x}_0 + \rmi \dot{y}_0 =
V_0(\bar z)|_{z=z_0} \end{equation} where in the complex velocity
\begin{eqnarray}\fl\bar V_0(z) = \frac{\rmi\kappa}{z (q-1)}\left[ Ln_q \left(1 - \frac{z}{z_0}  \right) - Ln_q \left(1 - \frac{z_0}{z}  \right)
+ Ln_q \left(1 - \frac{r^2_2}{z\bar z_0}  \right) - Ln_q \left(1 - \frac{z \bar z_0}{r^2_1}  \right)\right]\nonumber\\
 = \sum_{n= \pm 1}^{\pm\infty}\frac{\rmi \kappa}{z-z_0 q^n}- \sum_{n=
\pm 1}^{\pm\infty}\frac{\rmi \kappa}{z- \frac{r_1^2}{\bar z_0}
q^n},\nonumber
\end{eqnarray}
where $q = r^2_2/r^2_1$.
Contribution of the vortex on itself is excluded. If we take into account that q-harmonic series
\begin{equation} H(q) \equiv \sum_{n=1}^\infty \frac{1}{[n]} = - Ln_q 0\end{equation}
converges for $q > 1$, then at $z = z_0$ the first two terms cancel each other and we get the following equation of motion \begin{equation} \dot z_0 =
\frac{\rmi\kappa}{\bar z_0 (q-1)}\left[Ln_q \left(1 -
\frac{|z_0|^2}{r_1^2}  \right) - Ln_q \left(1 -
\frac{r_2^2}{|z_0|^2}  \right)\right].\label{vortexnonlinearoscillator}\end{equation}
Here to avoid manipulations with infinite sums we introduced q-logarithm function
\begin{equation} Ln_q(1- x)\equiv -\sum_{n=
1}^{\infty}\frac{x^n}{[n]},\,\,|x| < q, \,\,q>1,
\label{def2}\end{equation} where the q-number
\begin{equation}
[n] \equiv 1 + q + q^2 + ...+ q^{n-1} = \frac{q^{n}-1}{q-1}
\end{equation}
for any positive integer $n$.
Equation (\ref{vortexnonlinearoscillator}) is a nonlinear oscillator
\begin{equation}
\dot z_0 = - \rmi \omega z_0
\end{equation}
with frequency depending on amplitude  (and energy)
\begin{equation}
\omega(|z_0|^2) = \frac{\Gamma}{2\pi (q-1) |z_0|^2}\left[Ln_q \left(1 -
\frac{|z_0|^2}{r_1^2}  \right) - Ln_q \left(1 -
\frac{r_2^2}{|z_0|^2}  \right)\right].\end{equation}
In addition to the energy, another conserved quantity in this problem is an angular momentum
$
L = \Gamma \bar z_0 z_0$.
 In Hamiltonian form we have
 \begin{equation}
 \Gamma \dot z_0 = - 2\rmi \frac{\partial H}{\partial \bar z_0},
\end{equation}
with canonical bracket
\begin{equation}
\{ z_0, \bar z_0  \} = - \frac{2\rmi}{\Gamma},
\end{equation}
and Hamiltonian function
\begin{equation}
H(z_0, \bar z_0) = \frac{\Gamma^2}{4\pi} \ln \left| e_q \left( \frac{|z_0|^2}{(1-q) r^2_1}\right)   e_q\left( \frac{r^2_2}{(1-q) |z_0|^2}\right) \right|
,\end{equation}
where the Jackson q-exponential function is defines as
\begin{equation} e_q(z) = \sum_{n=0}^{\infty}
\frac{z^n}{[n]!}.\label{qexp1}\end{equation}
It is entire
in $z$ if $|q| > 1$, and admits infinite product representation
\begin{equation} \prod_{k=1}^\infty \left(1 -
\frac{z}{q^k}\right) = \frac{1}{1-z} e_q\left(\frac{-z}{1 -
q^{-1}}\right),\label{exprod}\end{equation}
showing that zeros of this function are ordered in geometric progression with ratio $q$.
   By introducing the action-angle variables $(J, \theta)$
   \begin{equation}
   z_0 = \rmi\sqrt{J} \rme^{-\rmi\theta},\,\,\,\,\,\,\bar z_0 = -\rmi\sqrt{J} \rme^{\rmi\theta},
   \end{equation}
   with canonical bracket
   \begin{equation}
  \{ \theta, J  \} = \frac{2}{\Gamma}
   \end{equation}
   we get the Hamiltonian function in terms of action variables $J$ only
   \begin{equation}
H(J) = \frac{\Gamma^2}{4\pi} \ln \left| e_q \left( \frac{J}{(1-q) r^2_1}\right)   e_q\left( \frac{r^2_2}{(1-q) J}\right) \right|\label{hamilton}
\end{equation}
and the frequency of rotation
\begin{equation}
\omega(J) = \frac{\partial H(J)}{\partial J} = \frac{\Gamma}{2\pi (q-1) J}\left[Ln_q \left(1 -
\frac{J}{r_1^2}  \right) - Ln_q \left(1 -
\frac{r_2^2}{J}  \right)\right].\label{frequency}\end{equation}
The angular momentum then is just
$
L = \Gamma J
$.
      To represent our model as an $f$- oscillator we introduce complex functions
      \begin{equation}
      z_f = \sqrt{\frac{H(z_0, \bar z_0)}{z_0 \bar z_0}} z_0,\,\,\,\,\,\bar z_f = \sqrt{\frac{H(z_0, \bar z_0)}{z_0 \bar z_0}} \bar z_0. \label{ftransformation}
      \end{equation}
Then we have simply f-oscillator
\begin{equation}
H(z_f, \bar z_f) = z_f \bar z_f,
\end{equation}
with Poisson bracket
\begin{equation}
\{ z_f, \bar z_f  \} = - \frac{2\rmi}{\Gamma} \frac{\partial H}{\partial J}(z_f, \bar z_f) = - \frac{2\rmi}{\Gamma} \omega(z_f, \bar z_f)
\end{equation}
and evolving with the  frequency (\ref{frequency})
\begin{equation}
\dot z_f = - \rmi \omega z_f.
\end{equation}

 \subsubsection{F-oscillator quantization of vortex motion}
In semiclassical approach, quantization is implemented by Bohr-Zommerfeld quantization rule of replacing  $J \rightarrow n+1/2$.
Then the energy spectrum is
  \begin{equation}
E_n = \frac{\Gamma^2}{4\pi} \ln \left| e_q \left( \frac{(n+\frac{1}{2})}{(1-q) r^2_1}\right)   e_q\left( \frac{r^2_2}{(1-q) (n+ \frac{1}{2})}\right) \right|.
\end{equation}

The    f-oscillator
    quantization of this system, replaces complex variables $z_0$ with bosonic operators $[a, a^+] = 1$, $N = a^+ a$.
Corresponding f- oscillator is given by (\ref{fHamilton}), (\ref{ftrans}) for Hamiltonian (\ref{hamilton}) H(N)
   and spectrum is
 \begin{equation}
 E_n = \frac{1}{2} [H(n) + H(n+1)].
 \end{equation}
In a similar way one can study N vortex polygon rotation in annular domain \cite{PY1} as a nonlinear or f-oscillator.

\section{Golden quantum oscillator}
As a final variation here we consider two golden ratio bases quantum q-oscillator.
We define creation and annihilation operators $b$ and $b^+$ in the Fock basis $\{ |n>\}$, $n = 0,1,2...$
represented by infinite matrices

\begin{eqnarray}
 b = \left( \begin{array}{cccc} 0  & \sqrt{F_1} & 0 & ...  \\
                                0 & 0 & \sqrt{F_2} & 0     \\
                                0 & 0 &  0 & \sqrt{F_3} \\
                                ... & ... & ... & ... \end{array}\right) ,\,\, b^+ = \left( \begin{array}{cccc} 0  & 0 & 0 & ...  \\
                                \sqrt{F_1} & 0 & 0 & ...     \\
                                0 & \sqrt{F_2} &  0 & ... \\
                                ... & ... & ... & ... \end{array}\right), \label{feq1}
\end{eqnarray}
where
$F_n$ are Fibonacci numbers. By introducing the Fibonacci operator as a matrix Binet formula,
\begin{equation}
F_N = \frac{\varphi^N - \varphi'^N}{\varphi - \varphi'},
\end{equation}
 where $N = a^+ a$ is the standard number operator, and $\varphi$, $\varphi'= - \varphi^{-1}$ are solutions of $\phi^2 = \phi + 1$, $\varphi= \frac{1 + \sqrt{5}}{2}$ - is the golden ratio, we find that in the Fock basis the eigenvalues are just Fibonacci numbers
\begin{equation}
F_N |n>= F_n  |n> \label{Fibnumbereigenvalue}
\end{equation}
and in the matrix form
\begin{eqnarray}
 F_N = \left( \begin{array}{cccc} F_0  & 0 & 0 & ...  \\
                                0 & F_1 & 0 & 0     \\
                                0 & 0 &  F_2 & 0 \\
                                ... & ... & ... & ... \end{array}\right) ,\,\,  F_{N+I} = \left( \begin{array}{cccc} F_1  & 0 & 0 & ...  \\
                                0 & F_2 & 0 & 0     \\
                                0 & 0 &  F_3 & 0 \\
                                ... & ... & ... & ... \end{array}\right). \label{fib1}
\end{eqnarray}
It satisfies Fibonacci recursion rule $F_{N-I} + F_N = F_{N+I} $. Then we have
\begin{equation}
b b^+ = F_{N + I},\,\,\,\,\,\,b^+ b = F_{N},
\end{equation}
and the commutator is
\begin{equation}
[b, b^+] = F_{N+I} - F_{N} = F_{N-I}.
\end{equation}
From definition of $F_N$ we get the matrix identity
\begin{equation}
\varphi^N = \varphi F_N + F_{N-I},
\end{equation}
where
\begin{eqnarray}
 \varphi^N = \left( \begin{array}{cccc} 1  & 0 & 0 & ...  \\
                                0 & \varphi & 0 & 0     \\
                                0 & 0 &  \varphi^2 & 0 \\
                                ... & ... & ... & ... \end{array}\right)\end{eqnarray}
and the following deformed commutation relations
 \begin{equation}
 b b^+ - \varphi b^+ b = \varphi'^N, \,\,\,\,\,\,\,\,\,    b b^+ - \varphi' b^+ b = \varphi^N.
 \end{equation}
 The Hamiltonian \cite{PN}
 \begin{equation}
 H = \frac{\hbar \omega}{2}( b b^+ + b^+ b) = \frac{\hbar \omega}{2} (F_N + F_{N+I}) = \frac{\hbar \omega}{2} F_{N+2I}
 \end{equation}
 is diagonal
 \begin{eqnarray}
 H = \left( \begin{array}{ccccc}  \frac{\hbar \omega}{2}F_2 & 0 & 0 & 0 & ...  \\
                                0 & \frac{\hbar \omega}{2} F_3 & 0 & 0  & ...   \\
                                0 & 0 &  \frac{\hbar \omega}{2} F_4 & 0 \\ 0 & 0 &  0 & \frac{\hbar \omega}{2} F_5\\
                                ... & ... & ... & ... \end{array}\right) =
                                \left( \begin{array}{ccccc}  \frac{\hbar \omega}{2} & 0 & 0 & 0 & ...  \\
                                0 & \hbar \omega & 0 & 0  & ...   \\
                                0 & 0 &  \frac{3\hbar \omega}{2} & 0 \\ 0 & 0 &  0 & \frac{5 \hbar \omega}{2}\\
                                ... & ... & ... & ... \end{array}\right)
                                \end{eqnarray}
and gives the energy spectrum as the Fibonacci sequence
 \begin{equation}
 E_n = \frac{\hbar \omega}{2} F_{n+2}.
 \end{equation}
 These energy levels satisfy the Fibonacci three term relations
 \begin{equation}
 E_{n+1} = E_{n} + E_{n-1}
 \end{equation}
 and the difference  between levels
 \begin{equation}
 \Delta E_n = E_{n+1} - E_{n} = \frac{\hbar \omega}{2} F_{n+1}
 \end{equation}
 is growing as Fibonacci sequence. Then the relative distance
 \begin{equation}
 \frac{\Delta E_n}{E_n} = \frac{F_{n+1}}{F_{n+2}}
 \end{equation}
 for asymptotic states $n \rightarrow \infty$
 is given just by the Golden ratio
 \begin{equation}
 lim_{n \rightarrow \infty}  \frac{\Delta E_n}{E_n} = \frac{1}{\varphi}.
 \end{equation}
 This behavior drastically differs from the harmonic oscillator and shows that there is no simple classical limit for this Golden oscillator.
 In fact, the Hamiltonian function for corresponding classical system becomes complex valued.
 By transformation
 \begin{equation}
 b = a \sqrt{\frac{F_N}{N}} = \sqrt{\frac{F_{N+I}}{N+I}} a,\,\,\,\,\,\,b^+ =\sqrt{\frac{F_{N}}{N}} a^+ = a^+ \sqrt{\frac{F_{N+I}}{N+I}}
 \end{equation}
 one can show that eigenstates
 \begin{equation}
 |n>_F = \frac{(b^+)^n}{\sqrt{F_n!}}|0>_F
 \end{equation}
  coincide with the Fock states $\{ |n> \}$.
 Then we have
 \begin{equation}
 b^+ |n>_F = \sqrt{F_{n+1}}\,|n+1>_F,\,\,\,\,\,b\, |n>_F = \sqrt{F_{n}}\,|n-1>_F.
 \end{equation}

\subsubsection{Golden coherent states}
 We define the golden coherent states as eigenstates
 \begin{equation}
 b \, |\beta>_F = \beta \,|\beta>_F.
 \end{equation}
 Expanding these states in the Fock space $|\beta> = \sum^\infty_{n=0} c_n |n>$ we find
 recurrence relation
 \begin{equation}
 c_{n+1} \sqrt{F_{n+1}} = c_n \, \beta,
 \end{equation}
 giving
 \begin{equation}
 c_n = \frac{\beta^n}{\sqrt{F_n!}} \, c_0.
 \end{equation}
 We fix $c_0$ by normalization condition $<\beta|\beta> =1$ so that
 \begin{equation}
 |c_0|^2 = \left(\sum^\infty_{n=0} \frac{|\beta|^{2n}}{F_n !}\right)^{-1} = \left( e_F^{|\beta|^2} \right)^{-1},
 \end{equation}
 where we have introduced the Fibonacci exponential function
 \begin{equation}
  e_F^{z} = \sum^\infty_{n=0} \frac{z^n}{F_n!},
  \end{equation}
  which as easy to see is entire function of $z$.
  As a result we get normalized coherent state
  \begin{equation}
  |\beta>_F = \left( e_F^{|\beta|^2} \right)^{-1/2} \sum^\infty_{n=0} \frac{\beta^n}{\sqrt{F_n!}}|n>_F,
  \end{equation}
  with the scalar product
  \begin{equation}
  _F<\alpha|\beta>_F = \frac{e^{\bar\alpha \beta}_F}{\left( e^{|\alpha|^2}_F\right)^{1/2} \left( e^{|\beta|^2}_F\right)^{1/2}}.
  \end{equation}

\subsubsection{Golden Fock-Bargman representation}
  For an arbitrary state from the Fock space $|\psi> = \sum^\infty_{n=0} c_n |n>_F$
  by projection
  \begin{equation}
  <\beta| \psi> = \left( e_F^{|\beta|^2}\right)^{-1/2}\sum^\infty_{n=0} c_n \frac{\bar \beta^n}{\sqrt{F_n!}}
  \end{equation}
  we find the analytic wave function
  \begin{equation}
  \psi (\beta) = \sum^\infty_{n=0} c_n \frac{\beta^n}{\sqrt{F_n!}}
  \end{equation}
  in the golden Fock-Bargman representation. By simple calculation it is easy to see that operators $b$ and $b^+$ in this
  representation are given by
  \begin{equation}
  b \rightarrow D^F_\beta, \,\,\,\, b^+ \rightarrow \beta,
  \end{equation}
  where the Binet-Fibonacci complex derivative we define as
  \begin{equation}
  D^F_z \psi(z) = \frac{\psi (\varphi z)- \psi(\varphi' z)}{(\varphi - \varphi') z} = \frac{\psi (\varphi z)- \psi(-\varphi^{-1} z)}{(\varphi + \varphi^{-1}) z}.
  \end{equation}
Action of this derivative on monomial gives just Fibonacci numbers
$D^F_z z^n = F_n z^{n-1}$
and for the Fibonacci exponential function we have
$
  D^F_z e_F^{z} = e_F^z$.
Then the Fibonacci number operator is represented as
\begin{equation}
  F_N \rightarrow  \beta D^F_\beta.
  \end{equation}
If an analytic function $\psi(z)$ is scale invariant $\psi_k(\lambda z) = \lambda^k \psi_k(z)$,
then it satisfies equation
\begin{equation}
D^F_z \psi_k(z) = \frac{\psi_k (\varphi z)- \psi_k(-\varphi^{-1} z)}{(\varphi + \varphi^{-1}) z} = \frac{ (\varphi)^k- (-\varphi^{-1})^k}{(\varphi + \varphi^{-1}) z} \psi_k(z)
\end{equation}
or
\begin{equation}
z D^F_z \psi_k(z) = F_k \,\psi_k(z).\label{fdifferenceequation}
\end{equation}
This eigenvalue problem is just the golden Fock-Bargman representation of the Fibonacci operator eigenvalue problem (\ref{Fibnumbereigenvalue}), where
eigenfunctions
\begin{equation}
\psi_k(z) = \frac{z^k}{\sqrt{F_k !}}
\end{equation}
are scale invariant.
However if we look for general solution of (\ref{fdifferenceequation}), then it is of the form
\begin{equation}
f_k (z) = z^k A(z)
\end{equation}
where $A(z)$ is an arbitrary golden-periodic analytic function $A(\varphi z) = A(-\varphi^{-1} z)$. Such a structure characterizes the quantum fractals \cite{Vitiello1},\cite{Pashaev2} and requires additional studies.

\section*{Appendix}

\appendix

\section{NLS hierarchy and q-NLS}
\subsection{NLS hierarchy}

We consider the NLS hierarchy
\begin{equation} \rmi\sigma_3  \left (\begin{array}{clcr}\psi \\
\bar\psi \end{array} \right)_{t_{N}}= {\cal{R}}^{N} \left (\begin{array}{clcr} \psi\\
\bar\psi \end{array} \right) \label{NLShierarchy}\end{equation}
where $t_N$, $N = 1, 2, 3, ...$ is an infinite time hierarchy.
Here ${\cal R}$ is the matrix integro-differential operator - the
recursion operator of the NLS hierarchy -
\begin{equation} {\cal{R}} = \rmi\sigma_3\left(\begin{array}{cccr}\partial_x+ 2\kappa^2 \psi
\int^x \bar\psi & -2\kappa^2 \psi \int^x \psi
\\ & \\-2\kappa^2 \bar\psi \int^x
\bar\psi&\partial_x+2\kappa^2\bar\psi \int^x \psi\end{array}
\right) \, \label{recursion}
\end{equation}
and $\sigma_3$ - the Pauli matrix.
For the first few members of the
hierarchy N = 1,2,3,4 this gives
\begin{equation}
\psi_{t_1} = \psi_x \, ,
\end{equation}

\begin{equation}
\rmi \psi_{t_2} + \psi_{xx} + 2\kappa^2|\psi|^2 \psi = 0 \, ,\label{NLS}
\end{equation}

\begin{equation}
 \psi_{t_3} + \psi_{xxx} + 6\kappa^2|\psi|^2 \psi_x = 0 \, ,
\end{equation}

\begin{equation}
 \rmi\psi_{t_4} = \psi_{xxxx} + 2\kappa^2\left(2 |\psi_x|^2 \psi + 4|\psi|^2 \psi_{xx}
 + \bar\psi_{xx}\psi^2 + 3 \bar\psi\psi^2_x
 \right) + 6 \kappa^4 |\psi|^4 \psi .
\end{equation}
In the linear approximation, when $\kappa = 0$, the recursion
operator is just the momentum operator
\begin{equation}
{\cal{R}}_0 = \rmi \sigma_3 \frac{\partial}{\partial x}
\label{reco}
\end{equation}
and the NLS hierarchy (\ref{NLShierarchy}) becomes the linear
Schrodinger hierarchy
\begin{equation}
\rmi\psi_{t_n} = \rmi^n \partial^n_x \psi \, .
\end{equation}
Written in the Madelung representation it produces the complex
Burgers hierarchy so that this representation plays the role of the
complex Cole-Hopf transformation
\cite{PG}.
Every equation of the hierarchy  (\ref{NLShierarchy}) is integrable.
The linear problem for the
$N$-th equation is given by the Zakharov-Shabat problem

\begin{equation}
\frac{\partial }{\partial x}\left(\begin{array}{c} v_1\\
v_2\end{array} \right) = \left( \begin{array}{cc} -
\frac{\rmi}{2}p& -\kappa^2 \bar\psi \\ \psi &\frac{\rmi}{2}p
\end{array}\right)\left(\begin{array}{c} v_1\\
v_2\end{array} \right) = J_1 \left(\begin{array}{c} v_1\\
v_2\end{array} \right) , \label{ZS11}\end{equation}
for the space
evolution, and
\begin{equation}
\frac{\partial }{\partial t_N}\left(\begin{array}{c} v_1\\
v_2\end{array} \right) = \left( \begin{array}{cc} - \rmi A_N&
-\kappa^2 \bar C_N \\ C_N & -\rmi A_N
\end{array}\right)\left(\begin{array}{c} v_1\\
v_2\end{array} \right) = J_{0_N} \left(\begin{array}{c} v_1\\
v_2\end{array} \right) , \label{ZSTN}\end{equation}
for the time part. Coefficient functions $C_N$ and $A_N$ are  \cite{Pashaev3},
\begin{equation}
\left( \begin{array}{c} C_N\\ \bar C_N
\end{array}\right) = \sum_{k=1}^{N} p^{N-k} {\cal{R}}^{k-1}\left( \begin{array}{c} \psi\\ \bar \psi
\end{array}\right) = (p^{N-1} + p^{N-2} {\cal{R}} + ... + {\cal{R}}^{N-1}) \left( \begin{array}{c} \psi\\ \bar \psi
\end{array}\right)
\end{equation}
To write this expression in a compact form, by analogy with
q-calculus it is convenient to introduce notation of nonsymmetric  q-number
operator
\begin{equation}
1 + q + q^2 + ... + q^{N-1} \equiv [N]_q,
\end{equation}
where $q$ is a linear operator. Hence with operator $q \equiv
{\cal{R}}/p$ we have the finite Laurent part in the spectral
parameter $p$
\begin{equation}
1 + \frac{{\cal{R}}}{p} + \left(\frac{{\cal{R}}}{p}\right)^2 + ...
+ \left(\frac{{\cal{R}}}{p}\right)^{N-1} \equiv [N]_{{\cal{R}}/p}.
\end{equation}
Then we have shortly
\begin{equation}
\left( \begin{array}{c} C_N\\ \bar C_N
\end{array}\right) = p^{N-1}[N]_{{\cal{R}}/p}\left( \begin{array}{c} \psi\\ \bar \psi
\end{array}\right).\label{CN}
\end{equation}
In a similar way
\begin{equation}
A_N = - \frac{p^N}{2} - \rmi\kappa^2 \left( \int^x \bar\psi, - \int^x
\psi \right) \left( \begin{array}{c} C_N\\ \bar C_N
\end{array}\right)
\end{equation}
and due to (\ref{CN})
\begin{equation}
A_N = - \frac{p^N}{2} - \rmi\kappa^2 p^{N-1}\left( \int^x \bar\psi, -
\int^x \psi \right) [N]_{{\cal{R}}/p}\left(
\begin{array}{c} \psi\\ \bar \psi
\end{array}\right).\label{AN}
\end{equation}
Equations (\ref{ZSTN}),(\ref{CN}) and (\ref{AN}) give the time part of the linear problem
(the Lax representation) for the N-th flow of NLS hierarchy (\ref{NLShierarchy}).

\subsection{General NLS hierarchy equation}

For the time $t$ determined by the formal series
\begin{equation}
\partial_t = {\sum^\infty_{N=0}} E_N \partial_{t_{N}}
\end{equation}
where $E_N$ are arbitrary constants, the general NLS hierarchy
equation is \cite{Pashaev3}
\begin{equation} \rmi\sigma_3  \left (\begin{array}{clcr}\psi \\
\bar\psi \end{array} \right)_{t}= \left(E_0 + E_1 {\cal{R}} + ... + E_N {\cal{R}}^{N} +...\right)\left (\begin{array}{clcr} \psi\\
\bar\psi \end{array} \right) \label{GNLShierarchy}\end{equation}

\subsubsection{Linear problem}

Integrability of this equation is associated with
the Zakharov-Shabat problem (\ref{ZS11}) and the time evolution
\begin{equation}
J_0 = \sum_{N=0}^\infty E_N J_{0_N} = \left( \begin{array}{cc} -
\rmi A& -\kappa^2 \bar C \\ C & -\rmi A
\end{array}\right),
\end{equation}
where
\begin{equation}
\left( \begin{array}{c} C\\ \bar C
\end{array}\right)=
\sum_{N=0}^\infty E_N \left( \begin{array}{c} C_N\\ \bar C_N
\end{array}\right) = \sum_{N=1}^\infty E_N p^{N-1}[N]_{{\cal{R}}/p}\left( \begin{array}{c} \psi\\ \bar \psi
\end{array}\right).\label{CH}
\end{equation}
In the last equation we have used that for $N=0$, $C_0 = 0$. Then
we have
\begin{equation}
A = \sum_{N=0}^\infty E_N A_N = - \frac{1}{2}\sum_{N=0}^\infty E_N
p^N - \rmi\kappa^2 \left( \int^x \bar\psi, - \int^x \psi \right)
\left(
\begin{array}{c} C\\ \bar C
\end{array}\right).\label{AH}
\end{equation}
\subsubsection{Integrable nonlinearization}

The above equation (\ref{GNLShierarchy}) gives integrable
nonlinear extension of a linear Schr\"dinger equation with general
analytic dispersion. Let us consider the classical particle
system with the energy-momentum relation
\begin{equation}
E = E(p) = E_0 + E_1 p + E_2 p^2 + ...\label{dispersion}
\end{equation}
Then the corresponding time-dependent Schr\"odinger wave equation
is
\begin{equation}
\rmi \hbar \frac{\partial}{\partial t}\psi = H\left(-\rmi \hbar
\frac{\partial}{\partial x}\right)\psi,\label{Schr}
\end{equation}
where the Hamiltonian operator results from the standard
substitution for momentum $p \rightarrow -\rmi \hbar
\frac{\partial}{\partial x}$ in the dispersion (\ref{dispersion}).
Equation (\ref{Schr}) together with its complex conjugate can be
rewritten as a system
\begin{equation}
\rmi \hbar \sigma_3 \frac{\partial}{\partial t}\left (\begin{array}{clcr}\psi \\
\bar\psi \end{array} \right) = H\left(-\rmi \hbar \sigma_3
\frac{\partial}{\partial x}\right)\left (\begin{array}{clcr}\psi \\
\bar\psi \end{array} \right).\label{MSchr}
\end{equation}
The momentum operator here is just the recursion operator
(\ref{reco}) in the linear approximation
${\cal{R}}_0 = \rmi
\sigma_3 \frac{\partial}{\partial x}$. Hence (\ref{MSchr}) can be
rewritten as the linear Schr\"odinger equation with an arbitrary
analytic dispersion

\begin{equation}
\rmi \hbar \sigma_3 \frac{\partial}{\partial t}\left (\begin{array}{clcr}\psi \\
\bar\psi \end{array} \right) = H\left({\cal{R}}_0\right)\left (\begin{array}{clcr}\psi \\
\bar\psi \end{array} \right)= \left(E_0 + E_1 {\cal{R}}_0 + E_2 {{\cal{R}}_0}^2... \right) \left (\begin{array}{clcr}\psi \\
\bar\psi \end{array} \right).\label{LSchr}
\end{equation}
Then the nonlinear integrable extension of this equation appears
as (\ref{GNLShierarchy}), which corresponds to the replacement
${\cal{R}}_0 \rightarrow {\cal{R}}$, ($\hbar = 1$), so that
\begin{equation} \rmi\sigma_3  \left (\begin{array}{clcr}\psi \\
\bar\psi \end{array} \right)_{t}= H\left(\cal{R}\right)\left (\begin{array}{clcr} \psi\\
\bar\psi \end{array} \right) \label{GNLShierarchy1}.\end{equation}
From this point of view the standard substitution for classical
momentum $p \rightarrow -\rmi \hbar \frac{\partial}{\partial x}$
or equivalently  $p \rightarrow -\rmi \hbar
\sigma_3\frac{\partial}{\partial x} = {\cal{R}}_0$ for the
equation in spinor form, gives quantization in the form of the
linear Schr\"odinger equation. While substitution $p \rightarrow
{\cal{R}}$ gives "nonlinear quantization" and the nonlinear
Schr\"odinger hierarchy equation.
\subsubsection{The Lax representation}

The related Lax representation for equation (\ref{GNLShierarchy1})
is given by (\ref{CH}), (\ref{AH}). Using definition of
q-derivative
\begin{equation}
D_q^{(\zeta)}f(\zeta) = \frac{f(q \zeta)-
f(\zeta)}{(q-1)\zeta}\label{qderivative}
\end{equation}
for operator $q = {\cal{R}}/p$ we have relation
\begin{equation}
D_{{\cal{R}}/p}^{(p)} \zeta^N = [N]_{{\cal{R}}/p}\, p^{N-1}.
\end{equation}
Then equation (\ref{CH}) can be rewritten as
\begin{equation}
\left( \begin{array}{c} C\\ \bar C
\end{array}\right)=
 \sum_{N=1}^\infty E_N \, p^{N-1}[N]_{{\cal{R}}/p}\left( \begin{array}{c} \psi\\ \bar \psi
\end{array}\right)= \sum_{N=1}^\infty E_N D_{{\cal{R}}/p}^{(p)} \, p^N\left( \begin{array}{c} \psi\\ \bar \psi
\end{array}\right)
\label{CDH1}
\end{equation}
or using linearity of (\ref{qderivative}) and dispersion
(\ref{dispersion})
\begin{equation}
\left( \begin{array}{c} C\\ \bar C
\end{array}\right)=
 D_{{\cal{R}}/p}^{(p)}\sum_{N=0}^\infty E_N \, p^N\left(
\begin{array}{c} \psi\\ \bar \psi
\end{array}\right)= D_{{\cal{R}}/p}^{(p)}\, E(p)\left(
\begin{array}{c} \psi\\ \bar \psi
\end{array}\right).
\label{CDH2}
\end{equation}
Due to definition (\ref{qderivative}) it gives simple formula
\begin{equation} \left( \begin{array}{c} C\\ \bar C
\end{array}\right)=
  \frac{E({\cal{R}})- E(p)}{{\cal{R}} - p}\left(
\begin{array}{c} \psi\\ \bar \psi
\end{array}\right), \label{CDH3}
\end{equation}
where \begin{equation}\frac{E({\cal{R}})- E(p)}{{\cal{R}} - p} =
E_1 + E_2 ({\cal{R}}+ p) + E_3 ({\cal{R}}^2 + {\cal R} p + p^2) +
...
\end{equation}
 Then for $A$ we obtain
\begin{equation}
A =  - \frac{1}{2}E(p) - \rmi\kappa^2 \left( \int^x \bar\psi, -
\int^x \psi \right) \frac{E({\cal{R}})- E(p)}{{\cal{R}} - p}\left(
\begin{array}{c} \psi\\ \bar \psi
\end{array}\right).\label{AH3}
\end{equation}
Equations (\ref{CDH3}),(\ref{AH3}) give the Lax representation of
the general integrable NLS hierarchy model (\ref{GNLShierarchy1}).
It is worth to note here that special form of the dispersion $E=
E(p)$ is fixed by physical problem. In \cite{Pashaev3} we have constructed semi-relativistic NLS equation.
In Section 4 as another application we discuss symmetric
 q-NLS equation.

\section{Wedge theorem}
Proof:  First we show that $\Im F_q(z)|_{\Gamma_1} = 0$, where $\Gamma_1: \{z =x + i 0 \}$ is the real axis. Substituting $z=x$ to  (\ref{wedge2})
and using identities for even powers
\begin{eqnarray}
q^2 &=& q^{-2(n-1)} = \bar q\,^{2(n-1)}, \\
q^4 &=& q^{-2(n-2)}=\bar q\,^{2(n-2)}, \\
...\\
q^{2k} &=& q^{-2(n-k)} = \bar q\,^{2(n-k)},\\
...\\
q^{2(n-1)} &=& q^{-2} = \bar q\,^{2},
\end{eqnarray}
we find
\begin{equation}F_q(z)|_{\Gamma_1} = \sum^{n-1}_{k=0} f(q^{2 k}x) + \sum^{n-1}_{k=0} \bar f(\bar q\,^{2(n- k)}x) =
\sum^{n-1}_{k=0} f(q^{2 k}x) + \sum^{n-1}_{k=0} \bar f(\bar q\,^{2 k}x) , \end{equation}
where in the second sum we have changed summation order by substitution $n-k \rightarrow k$. This shows that on the real line
the complex potential is pure real and imaginary part vanishes.

Now we show that $\Im F_q(z)|_{\Gamma_2} = 0$, where $\Gamma_2: \{z =x \, \rme^{\rmi \frac{\pi}{n}} = q \,x\}$ is the second boundary line.
Substituting we have
\begin{equation}F_q(z)|_{\Gamma_2} = \sum^{n-1}_{k=0} f(q^{2 k +1}x) + \sum^{n-1}_{k=0} \bar f(q\,^{2 k + 1)}x). \end{equation}
By identities for odd powers
\begin{eqnarray}
q &=& q^{-(2 n-1)} = \bar q\,^{2n-1}, \\
q^3 &=& q^{-(2n-3)}=\bar q\,^{2n-3}, \\
...\\
q^{2n-3} &=& q^{-3} = \bar q\,^{3},\\
q^{2n-1} &=& q^{-1} = \bar q,
\end{eqnarray}
and as follows
$q^{2k+1} = q^{-2n + 2k +1} = \bar q\,^{2n - 2k -1}$,
we rewrite the sum
\begin{equation}F_q(z)|_{\Gamma_2} = \sum^{n-1}_{k=0} f(q^{2 k +1}x) + \sum^{n-1}_{k=0} \bar f(q\,^{2 k + 1)}x). \end{equation}
After changing order of summation in the second sum and shifting summation index we finally get
\begin{equation}F_q(z)|_{\Gamma_2} = \sum^{n-1}_{k=0} f(q^{2 k +1}x) + \sum^{n-1}_{k=0} \bar f(\bar q\,^{2 k + 1)}x). \end{equation}
which shows that the stream function vanishes at the boundary. This completes proof of the wedge theorem.

\end{document}